\documentclass[prb,floatfix,showpacs,superscriptaddress,twocolumn]{revtex4-1}
\usepackage{eurosym}
\usepackage{amsmath}
\usepackage{amssymb}
\usepackage{amsfonts}
\usepackage{graphicx}
\usepackage{dcolumn}
\usepackage{bm}
\usepackage{color}
\usepackage{chngcntr}
\usepackage{changes}
\usepackage{hyperref}
\usepackage{ulem}

\DeclareGraphicsExtensions{.pdf,.png,.jpg,.eps}

\begin{document}

\title{Electron-acoustic-phonon interaction in core/shell Ge/Si and Si/Ge nanowires}
\author{Dar\'{\i}o G. Santiago-P\'{e}rez}
\affiliation{Universidad de Sancti Spiritus ``Jos\'{e} Mart\'{\i} P\'{e}rez", Ave. de los M\'artires 360, CP 62100, Sancti Spiritus, Cuba}
\affiliation{CLAF - Centro Latino-Americano de F\'{\i}sica, Avenida Venceslau Braz, 71, Fundos, 22290-140, Rio de Janeiro, RJ, Brasil}
\author{C. Trallero-Giner}
\affiliation{Department of Theoretical Physics, Havana University, Havana 10400, Cuba}
\author{G. E. Marques}
\affiliation{Departamento de F\'{\i}sica, Universidade Federal de S\~{a}o Carlos, 13.565-905 S\~{a}o Carlos, Brazil}

\begin{abstract}
General expressions for the electron- and hole-acoustic-phonon deformation potential Hamiltonians ($H_{E-DP}$) are derived for the case of Ge/Si and
Si/Ge core/shell nanowire structures (NWs) with circular cross section. Based on the short-range elastic continuum approach and on derived analytical results,
the spatial confinement effects on the phonon displacement vector, the phonon dispersion relation and the electron- and hole-phonon scattering amplitudes
are analyzed. It is shown that the acoustic displacement vector, phonon frequencies and $H_{E-DP}$ present mixed torsional, axial, and radial
components depending on the angular momentum quantum number and phonon wavevector under consideration. The treatment shows that bulk group velocities
of the constituent materials are renormalized due to the spatial confinement and intrinsic strain at the interface. The role of insulating shell on the
phonon dispersion and electron-phonon coupling in Ge/Si and Si/Ge NWs are discussed.
\end{abstract}

\date{\today }
\maketitle

\section{\label{Introduction}Introduction}

Based on enforcement of Si nanowires (NWs) linked to thermal conductivity,~\cite{Khitun1999181} photodetectors~\cite{Servati200764} and solar cells,~\cite{Tian2007,nl9034384,Allon} nowadays a notable effort has been addressed to study Si/Ge and Ge/Si core/shell semiconductor NWs.~\cite{Peng-Logan-APL-2010,doi:10.1021/nl203356h,doi:10.1021/nl103718a,doi:10.1021/cr400261y} These typical type-II band structures display high mobility~\cite{PhysRevLett.98.026801} and can be used in many applications.~\cite{PhysRevB.77.195325,Peng-Tang-Logan-JPCM-2011,Huang-Shouting-Li-APL-2011} It is also well established that the band gap at $\Gamma$-point of core/shell Si/Ge and Ge/Si nanostructures increases with decreasing radius, an effect directly linked to the spatial confinement and to the intrinsic strain at the interface,~\cite{Peng-Tang-Logan-JPCM-2011,PhysRevB.80.115322} produced in turn by the 4\% lattice mismatch between Si and Ge materials.~\cite{PhysRevLett.92.236805} Yet, more interestingly, in the nanoscale regime the Si/Ge and Ge/Si core/shell structures show a direct band gap at $\Gamma$-point.~\cite{PhysRevB.74.165314,PhysRevB.76.205327,Peng-Logan-APL-2010}

Besides the intrinsic strain, it is important to considered the role of spatial confinement on  the acoustic-phonon modes and on the electron-phonon interaction.
The acoustic phonon dispersion is strongly modified~\cite{Pokatilov2005168} when the radius of the quantum wire is of the order or smaller than the phonon wavelength. The confined acoustic phonon in such a nanostructure plays an important role on the carrier scattering rate, on the flow of electric current, and on the mobility or carrier transport. A suppression of the thermal conductivity in the core/shell Si/Ge NWs has been reported in Refs.~\onlinecite{PhysRevB.84.165415,PhysRevB.85.205439,doi:10.1063/1.4807389}. It has been shown that certain combinations of the core/shell cross-section modulation and  the acoustic mismatch allow to control the thermal flux. This result is, in principle, a promissory candidate for thermoelectric applications.
Thus, the reduction of the thermal conductivity and the characteristic of the carrier mobility in core/shell NWs are directly linked to the confinement effects on the phonon dispersion relation.~\cite{Thermal,Thermal2,Mobility,Mobility2,doi:10.1063/1.4807389} Also, it is important to remark that the core/shell wire structures are useful for optical applications~\cite{nl403341x,Mayer2013} and for quantum computing engineering with spin qubits.~\cite{Hu2011,nl501242b}

Several works have been devoted to obtain the acoustic phonon dispersion in wires and core/shell nanowires using both \textit{ab initio} calculations~\cite{abinitio1,abinitio2} and phenomenological continuum approaches (see Refs.~\onlinecite{Pokatilov2005168,Huang1,Loss2014, app5040728} and references there in). In addition, studies of electron-phonon interaction for the conduction band have been reported.~\cite{Hattori2010880,JAP_104_053716,PhysRevB.51.4695}
However, the electron-acoustic-phonon interaction in core-shell NWs has not been fully tackled. A phenomenological theory, allowing for the
evaluation of the electron-phonon Hamiltonian due to a deformation potential interaction, for cylindrical structures with arbitrary radii for both core and shell at the nanoscale regime, is a central issue for understanding the fundamental physics of many of the phenomena aforementioned. In the present work we study the electron-phonon interaction in Ge-core/Si-shell and Si-core/Ge-shell NWs in the framework of the continuum model and the $\vec{k}\cdot \vec{p}$ band theory.

The paper is organized as follows: in Sec.~\ref{Electron-phonon} we write down the general expression for the electron- and hole-phonon
deformation-potential Hamiltonians. For the conduction band we assume the $\Gamma_{1c}$ symmetry to be valid for Ge/Si and Si/Ge NWs grown along [110],
while for the holes we adopt the Bir-Pikus Hamiltonian (BPH) for states near the top of the valence bands with $\Gamma_{15v}$ symmetry. Sec.~\ref{Confined-Phonon} is devoted to a description of the elastic continuum model and the general basis of solutions for the phonon amplitudes. We make special emphasis on the phonon spectrum
calculations, the role of the spatial confinement effect, the symmetry of the space of solutions and on the comparison with the homogeneous wire limit. In Sec.~\ref{Scattering-Rate} we present detailed derivations of the electronic-acoustic phonon scattering rate for conduction and valence bands and of the influence of the core and shell radii for electrons and holes on the scattering amplitudes. We report our main results in Sec.~\ref{Conclusion}. Finally, in the Appendices we summarize the most relevant technical elements in the development of the present work.

\section{\label{Electron-phonon}Electron-acoustic-phonon interaction}

We consider typical core/shell cylindrical NWs with core radius $r_{c} $, shell thickness $\Delta =r_{s}-r_{c}$, and the $z-$axis
parallel to the growth direction [110]. We assume that all parameters involved in the present theoretical model are piece-wise functions of $r$, that is, we have assumed the parameters of the constituent materials to be isotropic.

In the occupation number representation, the Hamiltonian of the electrons interacting with the acoustic phonons can be expressed as~\cite{Madel}
\begin{equation}
H_{\mathrm{e-ph}}=\sum_{\alpha ^{\prime },\alpha}M_{\alpha ^{\prime
},\alpha }\left[ a_{j}^{\dagger }(\mathbf{k}_{z})+a_{j}(-\mathbf{k}_{z})
\right] c_{\alpha ^{\prime }}^{\dagger }c_{\alpha }\;,
\label{Hel-ph}
\end{equation}
where $a_{j}^{\dagger }(\mathbf{k}_{z})$ ($a_{j}(-\mathbf{k}_{z})$) denotes the phonon creation (annihilation) operator in the $j-$branch with wavevector $
\mathbf{k}_{z} (-\mathbf{k}_{z})$ and $c_{\alpha ^{\prime }}^{\dagger }$ ($c_{\alpha }$), the corresponding operator for electron in the electronic state $\alpha ^{\prime }$ ($\alpha$). Here, $M_{\alpha ^{\prime },\alpha }$ takes into account the electronic scattering event between the states $\alpha \rightarrow \alpha ^{\prime }$ by the interaction with an acoustic phonon. It is well known that in Si and Ge semiconductors the electron-phonon coupling can be determined using the short range deformation potential (DP) model.~\cite{Cardona} In a first approach, we develop a theory where this interaction is treated in the same way as in the bulk DP approach. Nevertheless, it has been reported that the DP constants are anisotropic and that depend on the spatial confinement (see Ref.~\onlinecite{nl9034384} and references there in). Furthermore, the DP mechanism can be treated as a perturbation to the band energies due to the lattice distortion; as a consequence, the electron-phonon coupling
depends on the electronic band structure.~\cite{Cardona} As we stated above, the Ge/Si and Si/Ge core/shell nanowires grown in the [110] direction show a direct band gap at $\Gamma$-point of the Brillouin zone,~\cite{PhysRevB.74.165314,Fazzio_295706,PhysRevB.76.205327,Peng-Tang-Logan-JPCM-2011}
hence, the conduction band minimum shows a $\Gamma_{1c}$ symmetry, while the top valence band has a $\Gamma_{15v}$ one, respectively.

\subsection{Conduction band}

Following the above discussion, the electron-phonon scattering amplitude probability can be written as
\begin{equation}
M_{\alpha _{e}^{\prime },\alpha _{e}}=\langle \Psi _{\alpha _{e}^{\prime
}}|a(\Gamma _{1c})\nabla \cdot \mathbf{u}|\Psi _{\alpha _{e}}\rangle \;,
\label{M_Matrixelem}
\end{equation}
where $a(\Gamma _{1c})$ is the volume deformation potential,~\cite{Cardona} $\mathbf{u}$ is the phonon displacement vector in the branch $j$, and $|\Psi_{\alpha _{e}}\rangle $ is the electron wave function for the core/shell NW.

\subsection{Valence band}
For the scattering amplitude, $M_{\alpha _{h}^{\prime },\alpha _{h}}$, of a hole in the valence band interacting with an acoustic phonon we have
\begin{equation}
M_{\alpha _{h}^{\prime },\alpha _{h}}=\langle \Psi _{\alpha _{h}^{\prime
}}|H_{BP}|\Psi _{\alpha _{h}}\rangle \;,
\label{M_Matrixelem_Hole}
\end{equation}
where $|\Psi _{\alpha _{h}}\rangle $ is the hole wave function in the NW and $H_{BP}$ is the Bir-Pikus Hamiltonian for the $J=3/2$ valence band states.~\cite{bir1974symmetry,Cardona} Assuming the zinc-blende symmetry, the $H_{BP}$ Hamiltonian in cylindrical coordinates and in the framework of the axial approximation, can be written as
\begin{multline}
H_{BP}=\left[ a(\Gamma _{15v})-\frac{1}{2}b(\Gamma _{15v})\left(
J_{z}^{2}-J^{2}/3\right) \right] \nabla \cdot \mathbf{u}+ \\
b(\Gamma _{15v})\left[ \frac{1}{2}J_{\mp }^{2}\mathcal{X}^{\pm }+\sqrt{2}
\{J_{\mp },J_{z}\}\mathcal{Y}^{\pm }\right. \\
\left. +\frac{3}{2}(J_{z}^{2}-J^{2}/3)\varepsilon _{zz}\right] \text{ },
\label{eq:Hbp}
\end{multline}
with $a(\Gamma _{15v})$ and $b(\Gamma _{15v})$ being the volume and shear deformation potentials for the highest energy at $\Gamma_{15v}$ valence band,~\cite{Note5} $\mathcal{X}^{\pm
}=e^{\pm 2i\theta }(\varepsilon _{rr}-\varepsilon _{\theta \theta }\pm
2i\varepsilon _{r\theta })$, $\mathcal{Y}^{\pm }=e^{\pm i\theta
}(\varepsilon _{rz}\pm i\varepsilon _{\theta z}),$ $\{J_{\mp },J_{z}\}=\frac{
1}{2}(J_{\mp }J_{z}+J_{z}J_{\mp }),$ $J_{\pm }=(J_{x}\pm iJ_{y})/\sqrt{2}$, and $J_{i}$ the Cartesian angular momentum operators for a particle with
spin 3/2 and $\varepsilon _{ij}$ the components of the stress tensor (see Appendix~\ref{Phonon} Eq.~(\ref{stresscomp})).

\section{\label{Confined-Phonon}Acoustic-phonon dispersion}

For an evaluation of the Hamiltonian~(\ref{Hel-ph}) and, in consequence, the matrix elements (\ref{M_Matrixelem}) and (\ref{M_Matrixelem_Hole}), it is
necessary to know the dependence of the phonon displacement $\mathbf{u}$ as well as the phonon frequencies on the core/shell spatial symmetry. In the
framework of elastic continuum approach, the equation of motion for the acoustic phonon modes takes the form~\cite{Born}
\begin{equation}
\rho \omega ^{2}\mathbf{u}-\nabla \cdot \mathbf{\sigma }=0\;,
\label{Emotion}
\end{equation}
with $\rho$ the mass density, $\omega $ the phonon frequency and $\mathbf{\sigma }$ the mechanical stress tensor. Following Hooke's law, $\mathbf{\sigma =C\cdot \varepsilon}$, with $\mathbf{\varepsilon }$ the strain tensor, $\mathbf{C}$ the elastic stiffness tensor  and the results being compiled in the Appendix \ref{Phonon}, the equation of motion for the acoustic phonon takes the form

\begin{equation}
\rho \omega ^{2}\mathbf{u}=\nabla (\rho v_{_{L}}^{2}\nabla \cdot \mathbf{u}
)+\nabla \times (\rho v_{_{T}}^{2}\nabla \times \mathbf{u)}\;.
\label{Emotion2}
\end{equation}
The solution of~(\ref{Emotion2}) consists of one longitudinal ($L$) $\mathbf{u}_{\mathbf{L}}$ and two transverse ($T$) $\mathbf{u}_{\mathbf{T}_{
\mathbf{1}}},\mathbf{u}_{\mathbf{T}_{\mathbf{2}}}$ fields, i.e. $\mathbf{u=u}_{\mathbf{L}}\mathbf{+u}_{\mathbf{T}_{\mathbf{1}}}\mathbf{+u}_{\mathbf{T}_{
\mathbf{2}}}$. Since the system is not homogeneous, in general $\mathbf{u}_{\mathbf{L}}$, $\mathbf{u}_{\mathbf{T}_{\mathbf{1}}}$, $\mathbf{u}_{\mathbf{T}
_{\mathbf{2}}}$ are coupled by the matching boundary conditions at the interface. Thus, the acoustic dispersion relations for the \textit{L} and
\textit{T} branches are not independent and the normal modes become a hybrid combination of $L$, $T_{1}$ and $T_{2}$ phonon vibrational motions.

It is important to remark that the equation of motion~(\ref{Emotion2}) for $r < r_c$, or $r_c < r < r_s$, corresponds to an isotropic model where an average velocity for the sound is assumed. An analysis of the phonon calculations and more general expressions including the anisotropy are presented in Appendix~\ref{Phonon}. It is shown there the phonon frequency calculations present a discrepancy of $10\%$ in comparison with the isotropic model.

In cylindrical geometry, the solution of Eq.~(\ref{Emotion2}) has full axial symmetry; hence, the displacement vector in cylindrical coordinates can be cast
as $\mathbf{u=}(u_{r},u_{\theta },u_{z})\exp i(n\theta +k_{z}z)$. In consequence, and following the method of solution described in Refs.~\onlinecite{CubaLibro,PhysRevB.91.075312}, one can derive a general basis of solutions for $(u_{r},u_{\theta },u_{z})$, namely

\begin{multline}
\left(
\begin{array}{c}
u_{r} \\
u_{\theta } \\
u_{z}
\end{array}
\right) =A_{_{L}}\left(
\begin{array}{c}
q_{_{L}}r_{c}f_{n}^{~\prime }(q_{_{L}}r) \\
i\frac{nr_{c}}{r}f_{n}(q_{_{L}}r) \\
ik_{z}r_{c}f_{n}(q_{_{L}}r)
\end{array}
\right) + \\
A_{_{T_{1}}}\left(
\begin{array}{c}
k_{z}r_{c}f_{n}^{~\prime }(q_{_{T}}r) \\
i\frac{nk_{z}r_{c}}{rq_{_{T}}}f_{n}(q_{_{T}}r) \\
-iq_{_{T}}r_{c}f_{n}(q_{_{T}}r),
\end{array}
\right) +A_{_{T_{2}}}\left(
\begin{array}{c}
\frac{nr_{c}}{r}f_{n}(q_{_{T}}r) \\
iq_{_{T}}r_{c}f_{n}^{~\prime }(q_{_{T}}r) \\
0
\end{array}
\right) \;,
\label{amplitude}
\end{multline}
where $n=0,\pm 1,\pm 2,...$ labels for the azimuthal motion, $k_{z}$ is the $z$-component of the phonon wavevector, and $q_{_{L}}(q_{_{T}})$ is given by

\begin{equation}
q_{_{L}}^{2}(q_{_{T}}^{2})=\frac{\omega ^{2}}{v_{_{L}}^{2}(v_{_{T}}^{2})}
-k_{z}^{2}\;.
\label{q}
\end{equation}

In Eq.~(\ref{amplitude}), if $x^{2}>0$ $(x^{2}<0)$ the function $f_{n}(x)$ is taken as Bessel $J_{n}$ (or Infeld $I_{n})$ for $0\leq r\leq r_{c}$ and as
linear combination of $J_{n}$ and Neumann $N_{n}$ functions of integer order $n$ (or combination of $I_{n}(x)$ and MacDonald $K_{n}(x)$)~\cite{Abramowitz}
for $r_{c}\leq r\leq r_{s}$. From (\ref{amplitude}), it is easy to check that $\nabla \cdot \mathbf{u}_{\mathbf{L}}=-(q_{_{L}}^{2} \pm k_{z}^{2})A_{_{L}}r_{c}f_{n}(q_{_{L}}r)e^{i(n\theta+k_{z}z)}$ (sign + for the Bessel functions and - for the modified Bessel functions) with $\nabla \cdot \mathbf{u}_{\mathbf{T}_{\mathbf{1}}}= \nabla
\cdot \mathbf{u}_{\mathbf{T}_{\mathbf{2}}}=0$ and $\nabla \times \mathbf{u}_{\mathbf{
L}}=0$, underlying the transverse and longitudinal character of the fields $\mathbf{u}_{\mathbf{T}_{\mathbf{1}}}$, $\mathbf{u}_{\mathbf{T}_{2}}$ and $
\mathbf{u}_{\mathbf{L}}$.

The eigenfrequencies of the phonon modes are obtained by imposing appropriate boundary conditions. As in the case of optical phonons, the strains at the interface play an important role on the phonon frequencies (see Ref.~\onlinecite{Singh2011}). For the acoustic  phonons, the effects of lattice mismatch between Ge and Si are taken into account through the continuity of the normal component of the stress tensor. We consider a free boundary at the shell surface, $\mathbf{\sigma }\cdot \mathbf{e}_{r}|_{r_{s}}=0$. Besides, the mechanical displacement and the normal component of the stress tensor should
be continuous at the core/shell interface, i.e. $\mathbf{u}|_{r_{c}^{-}} = \mathbf{u}|_{r_{c}^{+}}$ and $\mathbf{\sigma }\cdot \mathbf{e}
_{r}|_{r_{c}^{-}}=\mathbf{\sigma }\cdot \mathbf{e}_{r}|_{r_{c}^{+}}$. We point out that in the case of free standing homogeneous nanowires the basis of
solutions (\ref{amplitude}) match those reported in Ref.~\onlinecite{Hattori2010880}.

The calculation for the acoustical modes in NWs with cylindrical symmetry is a complicated task. In general, the phonon displacement $\mathbf{u}$ has all
three components ($\mathbf{u}_{\mathbf{T}_{\mathbf{1}}}$, $\mathbf{u}_{\mathbf{T}_{2}}$ and $\mathbf{u}_{\mathbf{L}}$), since none of the coefficients $A_{_{L}}$, $A_{_{T_{1}}}$ and $A_{_{T_{2}}}$ is zero, therefore, it cannot be decoupled into independent motions. Fixing $n$ and $k_{z}$, the constants $A_{_{L}}$, $A_{_{T_{1}}}$ and $A_{_{T_{2}}}$ are fully determined by the matching condition at $r=r_{c}$ and by the boundary condition of free standing NWs at $r=r_{s}$. Due to the cylindrical symmetry we cannot characterize the motions as pure torsional, dilatational or flexural modes. The resulting modes are  combination of
transverse and longitudinal characters. Nevertheless, from the symmetry of general basis~(\ref{amplitude}) we are able to obtain the following results: (i) for $n=0$ and $k_{z}=0$, we are in the presence of three independent $L$, $T_{1}$ and $T_{2}$ uncoupled modes with amplitudes $u_{r}(r)$, $u_{z}(r)$ and $u_{\theta }(r)$, respectively; (ii) for $n=0$ and $k_{z}\neq 0$, the longitudinal and transverse $T_{1}$ motions, $L-T_{1}$, are coupled, while $
T_{2}$ vibrational mode remains uncoupled; (iii) for $n\neq 0$ and $k_{z}=0$, the $T_{1}$ transverse phonon mode is independent, while the other two, $L$ and
$T_{2}$, are mixed; (iv) for $n\neq 0$ and $k_{z}\neq 0$ the longitudinal, the transverse $T_{1}$ and $T_{2}$ motions are coupled.
Below, we focus on the most relevant case of phonons with axial symmetry, $n=0$.

\subsection{Phonons with $k_{z}=0$}

\begin{figure}[htb]
\begin{center}
\includegraphics[width=\columnwidth]{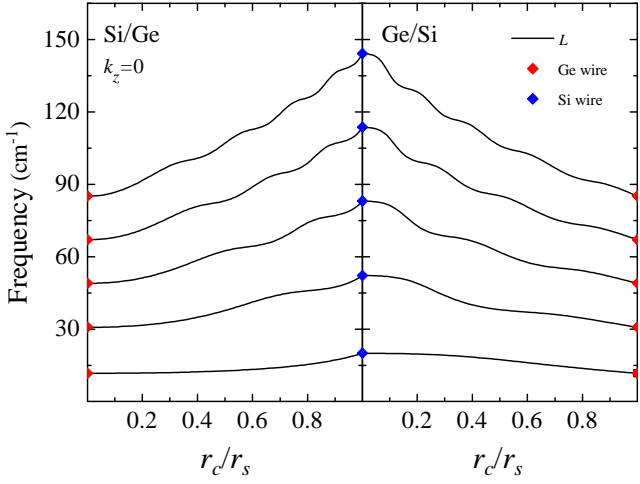}\\[0pt]
\end{center}
\par
\caption{(Color online) Frequencies of the first five breathing modes as a function of the ratio $r_{c}/r_{s}$ for fixed shell radius $r_{s}=5$ nm in
Si/Ge (left panel) and Ge/Si (right panel) NWs grown along the [110] crystallographic direction. The limits of Ge and Si nanowires are shown by
red and blue diamonds, respectively (see text).}
\label{bcont}
\end{figure}

\begin{figure}[tbp]
\begin{center}
\includegraphics[width=\columnwidth]{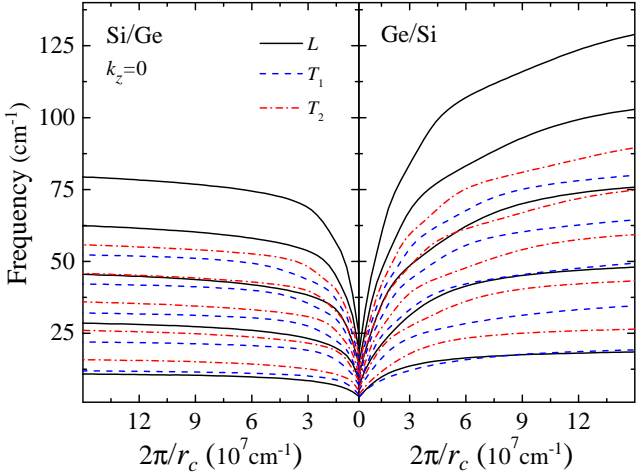}\\[0pt]
\end{center}
\par
\caption{(Color online) The same as Fig.~\ref{bcont} for the uncoupled $L$, $T_{1}$ and $T_{2}$ phonon modes as a function of $2\pi /r_{c}$, for fixed shell thickness $\Delta =r_{s}-r_{c}=5$ nm.}
\label{Deltacont}
\end{figure}
As stated above, the present case shows three uncoupled vibrations, $L$, $T_{1}$ and $T_{2}$. The longitudinal modes correspond to radial breathing mode (RBM) and their eigenfrequencies are ruled by the secular equation

\begin{multline}
F_{s}(\gamma \lambda _{_{L}}x/)[G_{s}(\lambda _{_{L}}x)J_{1}(x)-\rho
_{r}F_{c}(x)N_{1}(\lambda _{_{L}}x)]- \\
G_{s}(\gamma \lambda _{_{L}}x/)[F_{s}(\lambda _{_{L}}x)J_{1}(x)-\rho
_{r}F_{c}(x)J_{1}(\lambda _{_{L}}x)]=0\;,
\label{modosL}
\end{multline}
where $c$ $(s)$ labels the core (shell) region, $x=\omega r_{c}/v_{_{L_{c}}}$, $\lambda _{_{L}}=v_{_{L_{c}}}/v_{_{L_{s}}}$, $\gamma =r_{s}/r_{c}$, $\rho
_{r}=\rho _{s}/\rho _{c}$, $F_{i}(x)=v_{_{L_{i}}}^{2}xJ_{0}(x)-2v_{_{T_{i}}}^{2}J_{1}(x)$, and $G_{i}(x)=v_{_{L_{i}}}^{2}xN_{0}(x)-2v_{_{T_{i}}}^{2}N_{1}(x)$ ($i=s,c$). The RBM modes have been studied in the past for both nanotubes~\cite{Thomsen,Kurti2003} and semiconductor NWs.~\cite{Bourgeois,Trejo} Because of  their particular relevance it becomes necessary to focus on these modes in Ge/Si and Si/Ge core/shell nanowires. It is expected that the frequencies of the RBM modes, described by Eq.~(\ref{modosL}), are strongly dependent on the material composition, $\lambda _{_{L}}=v_{_{L_{c}}}/v_{_{L_{s}}}$ and size ratio, $\gamma =r_{s}/r_{c}$. Firstly, note from~(\ref{modosL}) that the two limiting cases $r_{c}=0$ and $r_{c}/r_{s}=1$ are given by the secular equation $F_{i}(z_{i})=0$ with $z_{i}=\omega r_{0}/v_{_{L_{i}}}$ ($i=s, c$) and $r_{0}$ the radius of the wire, i.e. the homogeneous NW dispersion relation is recovered for shell or core semiconductors.
Figure~\ref{bcont} shows the frequency dependence on the core/shell ratio $r_{c}/r_{s}$, with the limiting cases $r_{c}=0$ and $r_{c}/r_{s}=1$ shown by
diamonds. In the calculations we employed the following data for Si [Ge]: $v_{_{L}}=9.36$ $[5.39]\times 10^{5}cm/s$, $v_{_{T}}=5.25[3.30]\times 10^{5}cm/s$,~\cite{Note4}$\rho =2.33$ $[5.32]$ $g/cm^{3}$.~\cite{Adachi} The oscillations observed in Fig.~\ref{bcont} of $\omega$ as a function of the ratio $r_{c}/r_{s}$ can be explained by the interference between shell and core structures. Thus, for small values of $r_{c}/r_{s}$, the influence of the shell on the core phonon amplitude becomes stronger inhansing the oscillations. Moreover, the lower phonon frequencies are less affected showing almost a flat dispersion as a function of $r_{c}/r_{s}$, while the higher excited modes are more sensitive and displaying pronounced oscillations. The same trend is obtained for the $T_{1}$ and $T_{2}$ phonon modes (see Fig.~\ref{Deltacont}).

The confined eigenfrequencies, $\omega(k_{z}=0)$, for the $T_{2}$ modes can be obtained from the general expression
\begin{equation}
x_{s}J_{1}(x_{c})P_{22}(x_{s})+\frac{\rho _{r}}{\lambda _{_{T}}^{2}}
x_{c}J_{2}(x_{c})P_{12}(x_{s})=0\;.
\label{T2}
\end{equation}
Here, $x_{c}[x_{s}]=r_{c}\sqrt{(\omega /v_{_{T_c}}[v_{_{T_s}}])^{2}-k_{z}^{2}}$, $\lambda _{_T}=v_{_{T_{s}}}/v_{_{T_{c}}}$ and $P_{n,m}(x)=J_{n}(x)N_{m}( \gamma x)-J_{m}( \gamma x)N_{n}(x)$. Also, in the particular case when $k_{z}=0$, is possible to get an explicit expression for the $T_{1}$ frequency mode.

Figure~\ref{Deltacont} displays the dependence on $2\pi /r_{c}$ of the uncoupled $L$, $T_{1}$ and $T_{2}$ phonon frequency modes for fixed shell thickness $\Delta =$ $r_{s}-r_{c}$. In the limit $r_{c}\rightarrow \infty $ we recover the phonon frequencies for pure Si and Ge wires. As $r_{c} \rightarrow \infty$, we find that the phonon frequency resembles the typical linear acoustic bulk phonon dispersion as a function of the phonon wavevector. The spatial confinement renormalizes the sound velocity and we can rewrite, for large values of $r_{c}$, that $\omega _{_{L,T}}^{(j)}=(2\pi /r_{c})v_{_{L,T}}^{(j)}$ ($j=1, 2,...$) with different slope $v_{_{L,T}}^{(j)}$ for each mode. Notice that the cylindrical symmetry breaks the $T_{1}$ and $T_{2}$ degeneracy and two different sound velocities, $v_{_{T_{1}}}^{(j)}$ and $v_{_{T_{2}}}^{(j)}$, appear.

\subsection{Phonon dispersion with $k_{z}\neq0$}

\begin{figure}[tbp]
\begin{center}
\includegraphics[width=\columnwidth]{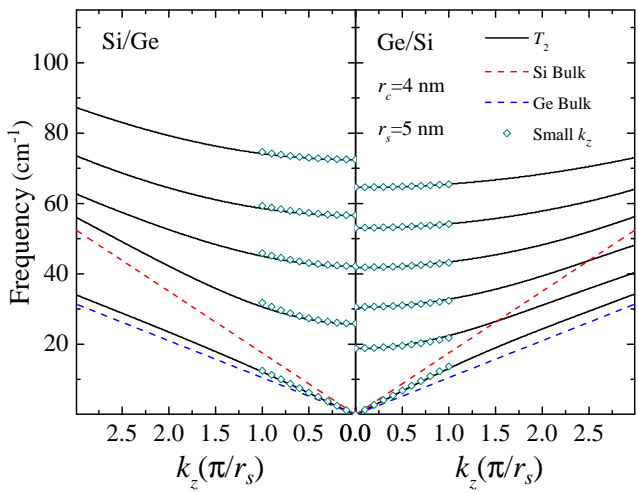}\\[0pt]
\end{center}
\par
\caption{(Color online) Phonon dispersion for uncoupled $T_{2}$ modes as a function of the phonon wavevector $k_{z}$ in units of $\pi /r_{s}$. Left panel Si/Ge; right panel Ge/Si NWs. Dashed lines represent the bulk dispersion relations for Si and Ge. Open diamonds are solutions of Eqs.~(\ref{wlineal}) and (\ref{APRT2}).}
\label{kz_pequeno}
\end{figure}

Following the secular Eq.~(\ref{T2}), in Fig.~\ref{kz_pequeno} we display the pure confined transverse $T_{2}$ phonon dispersion. For sake of
comparison the bulk phonon dispersions, $\omega _{_{\text{Ge}}}(k_{z})$ and $\omega _{_{\text{Si}}}(k_{z})$, are represented by blue and red dashed
lines, respectively. For small values of $k_{z}$ is possible to get useful analytical solutions. It should be remarked that in the case of Ge/Si,
if the set of the values ($\omega, k_{z})$ lies in the region $\omega _{_{\text{Ge}}}(k_{z})<\omega <\omega _{_{\text{Si}}}(k_{z})$, the parameter $x_{c}$ is
real while $x_{s}$ becomes a complex number. The opposite occur for  Si/Ge NWs. Accordingly, for the Ge/Si we get that the function $P_{nm}(x_{s})\Rightarrow P_{nm}(\left\vert x_{s}\right\vert)=$ $I_{n}(\left\vert x_{s}\right\vert )K_{m}( \gamma\left\vert x_{s}\right\vert )-I_{m}(\gamma\left\vert x_{s}\right\vert)K_{n}(\left\vert x_{s}\right\vert )$.

The numerical solution of Eq.~(\ref{T2}) shows a strong modification of the Si and Ge bulk phonon group velocities (see Fig.~\ref{kz_pequeno}) which
depend on the surrounding material, i.e. if the shell is composed by softer or harder material than the core semiconductor, the resulting group velocity
has lower or higher values. For example, in Ge/Si core/shell NWs the shell compresses the Ge core lattice while for Si/Ge the shell is compressed by the
core. Similar result have been achieved in Ref.~\onlinecite{Pokatilov2005168}.

Assuming small values of the wavenumber $k_z$, Eq.~(\ref{wlineal}) is obtained from Eq.~(\ref{T2}). Thus, the dispersion relation valid for Ge/Si ($x_c$ real, $x_s$ complex number) and Si/Ge ($x_c$ complex number, $x_s$ real) follows. This enables to better visualize the dependence of sound speed on materials parameters:
\begin{equation}
\omega =v_{_{T_{c}}}\sqrt{1+\frac{\left( \lambda _{_{T}}^{2}-1\right)
(\gamma ^{4}-1)\rho _{r}}{(\gamma ^{4}-1)\rho _{r}+1}}k_{z}=\overline{v_{_{T}}}k_{z}\;.
\label{wlineal}
\end{equation}
This equation shows that the lower modes present linear dependence in $k_{z}$, with a renormalized sound velocity $\overline{v_{_{T}}}$ that takes
into account the ratio between shell and core radii, as well as the densities and transverse velocities. Equation~(\ref{wlineal}) suggests the way to modify
the sound velocity as a function of the geometric factors ranging between the values ${v_{_{T}}^{\text{Ge}}}$ and ${v_{_{T}}^{\text{Si}}}$. The same expression has been found in Ref.~\onlinecite{Loss2014}.

In the domain ($\omega, k_{z})$ where $x_{c}$ and $x_{s}$ are both real functions, Eq.~(\ref{T2}) provides the dispersion relation for small values of $k_{z}$,
\begin{equation}
\omega _{_{T_{2}}}(k_{z})=\omega _{_{T_{2}}}(0)+\frac{1}{2}\frac{
v_{_{T_{s}}}^{2}}{\omega _{_{T_{2}}}^{2}(0)}k_{z}^{2}\;,
\label{APRT2}
\end{equation}
where $\omega _{_{T_{2}}}(0)$ is the confined phonon frequency of the core/shell NWs for $k_{z}=0$. In Fig.~\ref{kz_pequeno} the solutions given
by Eqs.~(\ref{wlineal}) and (\ref{APRT2}) are represented by open diamonds. By comparison with the numerical calculation of Eq.~\ref{T2}), it
can be seen that explicit expressions (\ref{wlineal}) and (\ref{APRT2}) are good approximations for $k_{z}(\pi /r_{s})\leq 1$.

\begin{figure}[tbh]
\begin{center}
\includegraphics[width=\columnwidth]{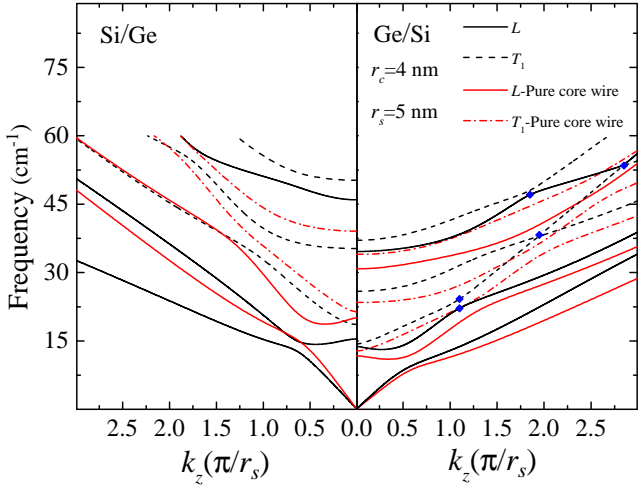}\\[0pt]
\end{center}
\caption{(Color online) The same as Fig.~\ref{kz_pequeno} for the mixed $L-T_1$ modes. Phonon dispersion relations for homogeneous Si and Ge
cylindrical wires are represented by red solid (longitudinal modes) and dashed-dots (transversal modes) lines. Full diamonds represent the anticrossings
between two nearby modes as explained in the text.}
\label{FigDesacoplados_kz}
\end{figure}

Another subset of solutions corresponds to the hybridized longitudinal and transverse motions. Fig.~\ref{FigDesacoplados_kz} shows phonon
dispersion of mixed $L-T_{1}$ modes for $\gamma =1.25$. The longitudinal ($L $) and transverse ($T_{1}$) labels are taken from the character
of the mode at $k_{z}=0$. For the sake of comparison, the phonon dispersions for the homogeneous Si and Ge cylindrical wires are shown in Fig.~\ref
{FigDesacoplados_kz}. Here, the corresponding longitudinal and transverse modes are represented by red solid and red dash-dots lines, respectively.
Due to the strain effect at the interface, it can be seen in the Fig.~\ref{FigDesacoplados_kz} that for the Ge/Si core/shell NW the phonon frequencies lye above the Ge wire, while the opposite is obtained for the Si/Ge NW, where $\omega _{\text{Si}}$ values are well above the core/shell Si/Ge phonon frequencies. At $k_{z}$
approaching zero, the lower mode presents a linear dependence of $\omega_{_{L-T_{1}}}$ on the wavenumber $k_{z}$ with certain effective sound velocity $
v_{_{L-T_{1}}}$ that depends on the radii $r_{c}$ and $r_{s}$.~\cite{Note2,Note3}
The bendings appearing in the Fig.~\ref{FigDesacoplados_kz} are manifestations of the strongest coupling between $L$ and $T_{1}$ modes. The mixed character of the states avoid crossing points in the phonon dispersion relation, i.e. the repulsion between near modes with the same symmetry occurs. This effect is observed in all dispersion relations having an important consequence in the electron-phonon Hamiltonian $H_{\mathrm{e-ph}}$ (see
discussion below). In Fig.~\ref{FigDesacoplados_kz}, some anticrossings associated with the mixing between $L$ and $T_{1}$ states, have been indicated by
full diamonds. The proximity of the levels belonging to the same space of solution or with the same symmetry is avoided by the repulsion between the
phonon states. At the anticrossings, a strong mixing between $L$ and $T_{1}$ states occurs and an exchange of character of the constants $A_{_{L}}$ and $
A_{_{T_{1}}}$ is obtained as a function of $k_{z}$.

Notice that higher excited states for $k_{z}\sim 0$ do not present strong mixing effect and the phonon dispersion relation can be described by simple parabolic law, $\omega_{_{L(T_{1})}}(k_{z})=\omega_{_{L(T_{1})}}(0)+\beta_{_{L(T_{1})}}^{2}k_{z}^{2}$. Here, $\beta_{_{L(T_{1})}}$ measures the curvature of the phonon dispersion and $\omega_{_{L-T_{1}}}(0)$ are the NW phonon frequencies for $k_{z}=0$. We arrived to the same results for the homogeneous Si and Ge cylindrical wires.

\section{\label{Scattering-Rate}Scattering rate}

Due to translational and cylindrical symmetries, the matrix element $M_{\alpha ^{\prime },\alpha }$ can be cast as follows

\begin{equation}
M_{\alpha ^{\prime },\alpha }=S_{e-ph}\delta_{{m}^{\prime},m+n}\delta_{{k}
^{\prime},k+k_{z}}\;,
\end{equation}
where the angular momentum and linear momentum conservations are written explicitly. $S_{e-ph}=\left\langle m^{\prime }\right\vert H_{e-ph}\left\vert m\right\rangle$ is the scattering amplitude due to the electronic transition assisted by an acoustical phonon, between the electron or hole states $\left\vert
m^{\prime}\right\rangle \rightarrow $ $\left\vert m\right\rangle $ (see Appendices~\ref{Conduction-Band} and \ref{Valence-Band}). For the phonon eigenvectors {$\mathbf{u}_{n,k_{z}}$} we choose the normalization condition

\begin{equation}
\int \rho (r)|\mathbf{u}_{n,k_{z}}(\mathbf{r})|^{2}dV=\frac{\hbar }{2\omega
_{n}(k_{z})}\;,
\label{Norma}
\end{equation}
with $\omega _{n}(k_{z})$ the acoustic-phonon dispersion of the core/shell problem. Let us discuss a general formulation for the electron-acoustic
deformation potential Hamiltonian, $H_{E-P}$ and an evaluation of the scattering amplitudes for the electrons and holes.

\subsection{Electron-LA Hamiltonian}

According to Eq.~(\ref{M_Matrixelem}), a transverse or torsional mode does not induce volume change and only the longitudinal acoustic motion $\mathbf{u
}_{\mathbf{L}}(\mathbf{r})$ contributes to electron-phonon Hamiltonian $H_{E-DP}$. Hence, by assuming $A_{_{L}}$ as independent constant in Eq.~(\ref{amplitude}), we have
\begin{multline}
H_{E-DP}=a(\Gamma _{1c})\nabla \cdot \mathbf{u}_{\mathbf{L}}= \\
-\sqrt{\frac{\hbar \omega _{n}^{3}(k_{z})}{4\pi r_{c}^{2}L\rho
_{c}v_{_{L}}^{4}}}\frac{a(\Gamma _{1c})}{\mathcal{N}_{n,k_{z}}}
f_{n}(q_{_{L}}r)e^{i(n\theta +k_{z}z)}\;,
\label{EDP}
\end{multline}
\begin{figure}[tbh]
\begin{center}
\includegraphics[width=\columnwidth]{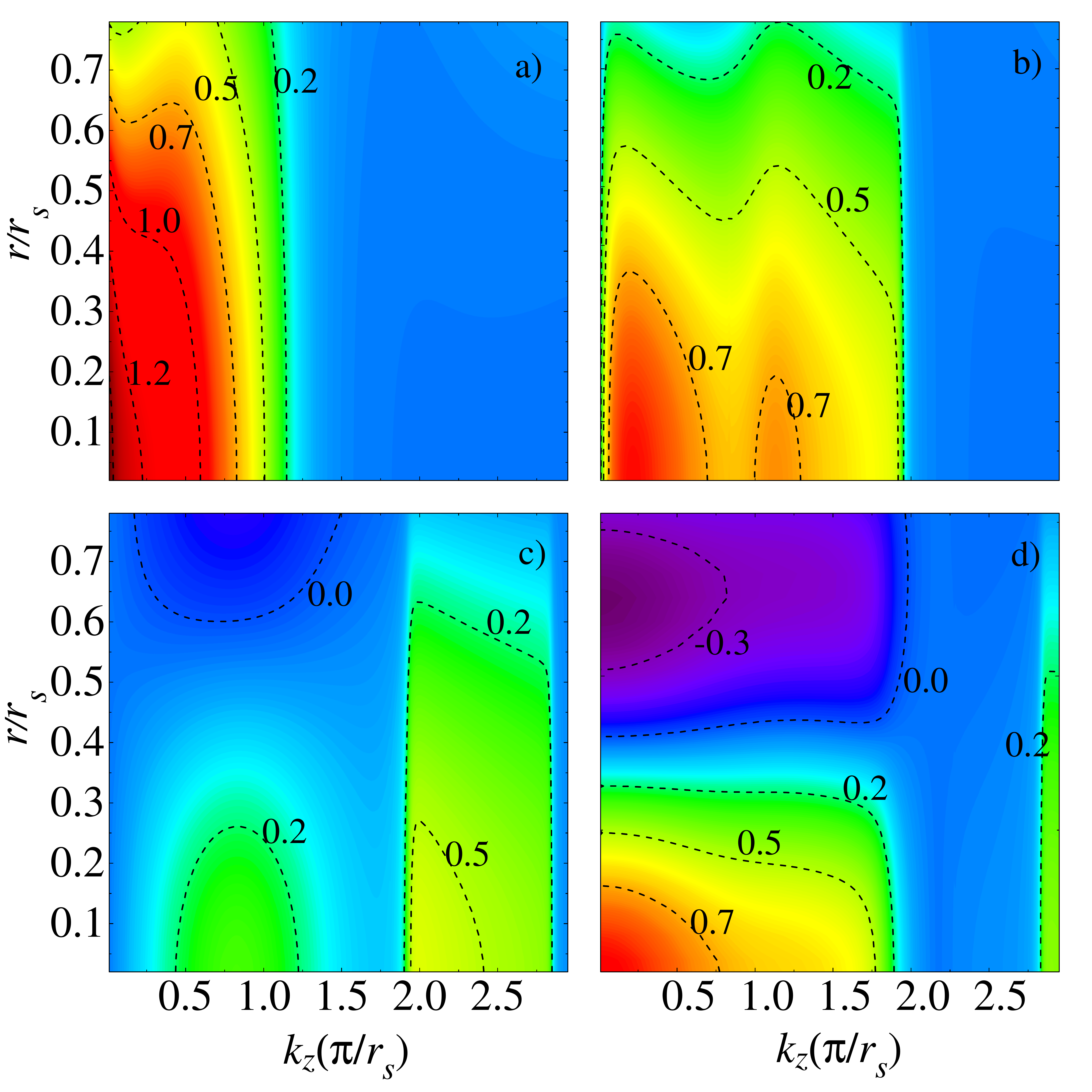}\\[0pt]
\end{center}
\caption{(Color online) Contour plots of the Hamiltonian (\ref{EDP}) in units of $H_{OE}$ as function dimensionless wavenumber $k_{z}/(\pi /r_{s})$ and radius $r/r_{c}$ for the phonon modes a) $\omega_{_{L}}^{(1)}$, b) $\omega _{_{T_1}}^{(1)}$, c) $\omega_{_{T_1}}^{(2)}$, and d) $\omega _{L}^{(2)}$ of the Ge/Si core/shell
NW (see text). In the calculation we fixed $k_{z}=0$ and $z=0$.}
\label{Fig5}
\end{figure}where $\mathcal{N}_{n,k_{z}}=\sqrt{\int_{0}^{\gamma }\rho (z)|\overline{\mathbf{u}_{n,k_{z}}(z)}|^{2}zdz/(\rho _{c}A_{_{L_{c}}}^{2})}$ is the
normalization constant for the dimensionless phonon amplitude $|\overline{\mathbf{u}_{n,k_{z}}(z)}|$. In Fig.~\ref{Fig5} the characteristic contour map for the electron-LA Hamiltonian~(\ref{EDP}) is shown for the Ge/Si NWs. We choose the first four modes of Fig.~\ref{FigDesacoplados_kz} where $\omega _{n=0}(k_{z}=0)\neq 0$. According to the general basis of solutions~(\ref{amplitude}), for $n=0$, the longitudinal displacement vector $\mathbf{u}_{\mathbf{L}}$ have non-zero radial and axial components. Figures~\ref{Fig5} a), b), c) and d) correspond to the uncoupled confined frequencies $\omega_{_{L}}^{(1)}$, $\omega_{_{T_{1}}}^{(1)}$, $\omega_{_{T_{1}}}^{(2)}$, and $\omega_{_{L}}^{(2)}$ for $k_{z}=0$ of the $L$ and $ T_{1}$ motions. In the panels a) and d) one finds, for $r=0$, the stronger spatial localization in correspondence with the longitudinal character of these modes. In panel a) we observe, as $k_{z}$ increases, that the $T_{1}$ component becomes stronger, in particular for $k_{z}>1.1$ the contribution of the $\omega _{_{L}}^{(1)}$ mode to $H_{E-DP}$ is almost zero. The same is observed in panel d) for the state $L^{(2)}$, but the limiting value is $k_{z}>1.8$. These two values of $k_{z}$ are in correspondence with the anticrossings shown by full diamonds in the Figure~\ref{FigDesacoplados_kz} for the $L^{(1)}$ and $L^{(2)}$ phonon states. Note in Fig.~\ref{FigDesacoplados_kz} that for $k_{z}\sim 2.8$ an anticrossing occurs and the $L^{(2)}$ mode presents stronger $L$ character and, in consequence, the spatial distribution of $H_{E-DP}$ is enhanced. In panel b) the observed strong spatial localization of $H_{E-DP}$ at $r=0$ with $k_{z}\approx 0.2$ is explained by the reduction of the coefficient $A_{_{T_{1}}}$ as a function of $k_{z}$, in the state $T_{1}^{(1)}$. Due to the anticrossing between the states $T_{1}^{(1)}$ and $T_{1}^{(2)}$ for $
k_{z}\approx 2$, the $\omega _{_{T_{1}}}^{(1)}$ mode is almost transverse and its contribution to the spatial distribution $f_{n=0}(q_{_{L}}r)|_{\omega
=\omega_{_{T_{1}}}^{(1)}}$ decays to zero. At the same time, the mode $T_{1}^{(2)}$ increases the $A_{L}$ amplitude, and $H_{E-DP}$ at $\omega
=\omega_{_{T_{1}}}^{(2)}$ increases in the region $1.9 < k_{z} < 2.9$ as seen in Fig.~\ref{Fig5}c). Thus, the phonon dispersion of the Ge/Si NWs for a given ratio of $r_{s}/r_{c}$ has a preponderant influence on the spatial distribution of $H_{E-DP}$ as a function of $k_{z}$.
\begin{figure}[tbh]
\begin{center}
\includegraphics[width=\columnwidth]{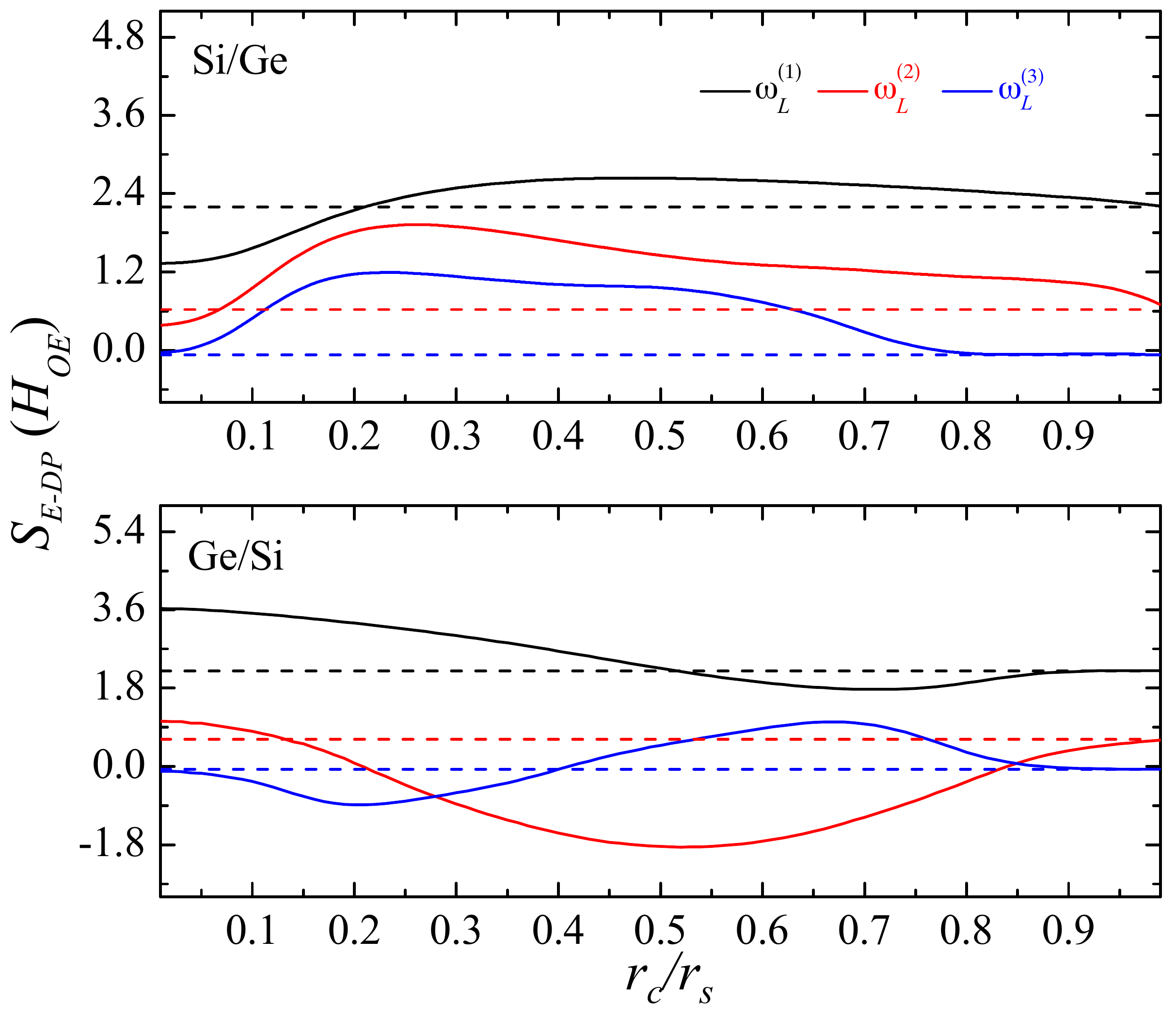}\\[0pt]
\end{center}
\caption{(Color online) Electron scattering amplitude for the excited frequencies $\omega_{_{L}}^{(1)}$, $\omega_{_{L}}^{(2)}$ and $\omega_{_{L}}^{(3)}$ (see text) as a function of the ratio $r_{c}/r_{s}$. Dashed lines: homogeneous wires as given by Eq.~(\ref{SHED}). Upper panel Si/Ge, lower panel Ge/Si.}
\label{Fig6}
\end{figure}
Taking into account the Eqs.~(\ref{EDP}), (\ref{electrwave}) and (\ref{ME}), the electron scattering amplitude can be written as
\begin{equation}
S_{E-DP}=\frac{1}{r_{c}^{2}}\left\langle F_{m_{e}^{\prime }}\right\vert
H_{OE}\frac{\omega _{n}}{v_{_{L}}\mathcal{N}_{n,k_{z}}}f_{n}\left\vert
F_{m_{e}}\right\rangle \;,
\label{SED}
\end{equation}
where $H_{OE}=-a(\Gamma _{1c})\sqrt{\hbar \omega _{n=0}(k_{z}=0)/4\pi r_{c}^{2}L_{0}\rho _{c}v_{_{L}}^{2}}$. From Eq.~(\ref{SED}) immediately follows that phonon modes with $n=0$ assist to electron intrasubband transitions, $m_{e}^{\prime }=m_{e}$, while for $n\neq 0$ intersubband transitions with $m_{e}^{\prime }\neq m_{e}$ occur fo. Note that in the case of homogeneous wire, $\left\langle F_{m_{e}^{\prime}}\right\vert f_{n}\left\vert F_{m_{e}}\right\rangle /r_{c}^{2}$ corresponds to the electron form factor or overlap integral between the normalized radial electronic states and the phonon function $f_{n}$ of the quantum wire. For comparison, we consider the Eq.~(\ref{SED}) for a homogeneous wire. Assuming the size-quantum limit (strong spatial confinement) where electrons populate the lowest subband $(m_{e}^{\prime}=m_{e}=0$ and $n=0)$ and intersubband transitions $|p_{e}^{\prime}\rangle \rightarrow |p_{e}\rangle $ are discarded, the scattering amplitude at $k_{z}=0$ reduces to

\begin{multline}
S_{E-DP}^{H}=H_{OE}\left( \frac{r_{c}^{2}q_{_{L}}^{2}}{4}\delta ^{4}-\delta
^{2}+1\right) ^{-\frac{1}{2}}
\label{SHED} \\
\times \frac{\langle J_{0}(p_{e})|J_{0}(q_{_{L}})|J_{0}(p_{e})\rangle }{
r_{c}^{2}J_{1}^{2}(p_{e}r_{c})J_{0}(q_{_{L}}r_{c})}\;,
\end{multline}
with $\delta =v_{_{L}}/v_{_{T}}$.

It is instructive to compare the behavior of the electron scattering amplitudes for core/shell Si/Ge and Ge/Si NWs. Fig.~\ref{Fig6} displays the reduced scattering amplitude $S_{E-DP}/H_{OE}$ as a function of the ratio $r_{c}/r_{s}$ for both core/shell NWs. In the calculation for the Si/Ge [Ge/Si] NWs we fixed the value of $H_{OE}$ with the parameters of Si [Ge] semiconductor. For each structure, in the quantum limit approach, the first three $L$ modes of the structure with frequencies $\omega_{_{L}}^{(j)}$ ($j=1,2,3 $)$\neq 0$ at $k_{z}=0$ are considered. In the figure the form factor, using Eq.~(\ref{SHED}) and $r_{c}=5$ nm, is represented by dashed lines. In  Si/Ge NWs, electrons are confined in the core whereas in Ge/Si NWs they are in the shell. For the evaluation of Eq.~(\ref{SED}) we employed the results displayed in Appendix~\ref{Conduction-Band}. In the upper panel of Fig.~\ref{Fig6} (Si/Ge NWs) the influence of the shell of Ge on the $S_{E-DP}$ is shown.  If $r_{c}=r_{s}$ we have a quantum wire of Si and $S_{E-DP}/H_{OE}$ as described by Eq.~(\ref{SHED}). If $r_{c}/r_{s}\neq 1$ we are in the presence of a Si/Ge core/shell system. Thus, we can observe that the value of $S_{E-DP}$, for the $\omega_{_{L}}^{(1)}$ modes, firstly increases, reaching a maximum for $\gamma _{\max }^{(1)}=(r_{c}/r_{s})_{\max }^{(1)}\approx 0.4$ and for $\gamma<\gamma _{\max }^{(1)}$ the quantity $S_{E-DP}/H_{OE}$ reaches asymptotically the homogeneous Si wire value. In the case of $\omega_{_{L}}^{(j)}$ ($ j=2,3 $), the reduced scattering amplitude grows, reaching a maximum value near ($r_{c}/r_{s})_{\max }^{(2)}\approx 0.23$. For $\gamma
<\gamma _{\max }^{(2)}$, $S_{E-DP}/H_{OE}$ decreases to the limiting value of Eq.~(\ref{SHED}). In the lower panel of Fig.~\ref{Fig6} (Ge/Si NWs) the wire of Ge is reached at $r_{c}=r_{s}$. From the figure we can observe the strong influence of the shell on the $S_{E-DP}^{H}$ for $r_{c}/r_{s} < 0.8$ besides oscillations of $S_{E-DP}$ around the $S_{E-DP}^{H}$ values, a fact reflecting the oscillatory behavior of the phonon modes with $\gamma $ (see Fig.~\ref{bcont}). A similar result for the electron scattering amplitude has been reported in Ref.~\onlinecite{Ramayya} for Si nanowires.

\subsection{Hole-Acoustical-Phonon Hamiltonian}

By employing the solutions for the phonon amplitudes~(\ref{amplitude}), the matrix representation of the angular momentum $J=3/2$~\cite{PhysRev.102.1030,PhysRevB.84.195314} and the strain relations given in Appendix~\ref{Phonon}, the hole scattering amplitudes for the Hamiltonian~(\ref{eq:Hbp}) can be cast as

\begin{equation}
S_{H-BP}=\left\langle \widehat{F}_{m_{h}^{\prime }}^{(i)}\right\vert
H_{BP}\left\vert \widehat{F}_{m_{h}}^{(i)}\right\rangle \;,
\label{SHbp}
\end{equation}
where $i=hh^{+}$, $lh^{+}$, $lh^{-},hh^{-}$,

\begin{widetext}
\begin{equation}
S_{H-BP}=\left\langle
\begin{array}{c}
a_{1i}F_{m_{h}+n} \\
a_{2i}F_{m_{h}+n+1} \\
a_{3i}F_{m_{h}+n+2} \\
a_{4i}F_{m_{h}+n+3}
\end{array}
\right\vert ^{\dagger }\left(
\begin{array}{cccc}
\mathcal{T}_{+} & \mathcal{Y}^{-} & \mathcal{X}^{-} & 0 \\
\mathcal{Y}^{+} & \mathcal{T}_{-} & 0 & \mathcal{X}^{-} \\
\mathcal{X}^{+} & 0 & \mathcal{T}_{-} & -\mathcal{Y}^{-} \\
0 & \mathcal{X}^{+} & -\mathcal{Y}^{+} & \mathcal{T}_{+} \\
\end{array}
\right) \left\vert
\begin{array}{c}
a_{1i}F_{m_{h}} \\
a_{2i}F_{m_{h}+1} \\
a_{3i}F_{m_{h}+2} \\
a_{4i}F_{m_{h}+3}
\end{array}
\right\rangle \;,
\label{SHHbp}
\end{equation}
\end{widetext}
\begin{widetext}
\begin{eqnarray}
\mathcal{T}_{\pm } &=&-\left[ A_{_{L}}\left( \left[ a(\Gamma _{15v})\pm
\frac{1}{2}b(\Gamma _{15v})\right] \frac{\omega ^{2}}{v_{_{L}}^{2}}\mp \frac{
3}{2}k_{z}^{2}b(\Gamma _{15v})\right) f_{n}(q_{_{L}}r)\pm \frac{3}{2}
A_{_{T_1}}b(\Gamma _{15v})k_{z}q_{_{T}}f_{n}(q_{_{T}}r)\right] \text{ },
\notag \\
\mathcal{Y}^{\pm } &=&\mp i\sqrt{3}b(\Gamma _{15v})\left[
A_{_{L}}k_{z}q_{_{L}}f_{n\pm 1}(q_{_{L}}r)\mp \frac{1}{2}\left[
A_{_{T_1}}\left( q_{_{T}}^{2}-k_{z}^{2}\right) +A_{_{T_2}}k_{z}q_{_{T}}\right]
f_{n\pm 1}(q_{_{T}}r)\right] \text{ },  \notag \\
\mathcal{X}^{\pm } &=&\frac{\sqrt{3}}{2}b(\Gamma _{15v})\left[
A_{_{L}}q_{_{L}}^{2}f_{n\pm
2}(q_{_{L}}r)+(A_{_{T_1}}k_{z}q_{_{T}}-A_{_{T_2}}q_{_{T}}^{2})f_{n\pm
2}(q_{_{T}}r)\right] \text{ }.
\label{MEL}
\end{eqnarray}
\end{widetext}

From Eqs.~(\ref{SHbp}), (\ref{MEL}) and the basis of solutions~(\ref{amplitude}), we extract the following conclusions: a) For phonon states with $n=0$ and $k_{z}=0$ we have three independent hole-phonon interaction Hamiltonians, accounting for the three uncoupled subspaces, $L$, $T_{1}$, $T_{2}$, with eigenfrequencies $\omega_{_{L}}$, $\omega_{_{T_{1}}}$ and $\omega_{_{T_{2}}}$, respectively. Evaluating~(\ref{MEL}) at $\omega =\omega _{_{L}}$ and using the fact that $A_{_{L}}\neq 0$ and $A_{_{T_{1}}}, A_{_{T_{2}}}=0$, we see that the Hamiltonian~(\ref{SHbp}) couples the diagonal intraband hole sates $\left\vert i\right\rangle
\Rightarrow \left\vert i\right\rangle $ and the weak coupling interband between $\left\vert lh^{\pm }\right\rangle \Leftrightarrow \left\vert hh^{\mp }\right\rangle $ states; if we choose $\omega =\omega _{_{T_{1}}}$ where $A_{_{L}},A_{_{T_{2}}}=0$ and $A_{_{T_{1}}}\neq 0$, we are in the presence of interband transitions $\left\vert lh^{\pm }\right\rangle \Leftrightarrow \left\vert hh^{\pm }\right\rangle$. Also, for $\omega =\omega_{_{T_{2}}}$
with $A_{_{L}},A_{_{T_{1}}}=0$ and $A_{_{T_{2}}}\neq 0$, results in scattering $ \left\vert lh^{\mp }\right\rangle \Leftrightarrow \left\vert hh^{\pm
}\right\rangle $. b) Fixing $n=0$ and $k_{z}\neq 0$ there are two independent subspaces, $L-T_{1}$ and $T_{2}$. The first one couples $L$ and $T_{1}$ motions, while the second corresponds to pure $T_{2}$ transverse phonons. Similar expressions are obtained for homogeneous wires by choosing properly the function $F_{m_{h}}(r)$ and $f_{n}(r)$ inside the cylinder.
\begin{figure}[tbh]
\begin{center}
\includegraphics[width=\columnwidth]{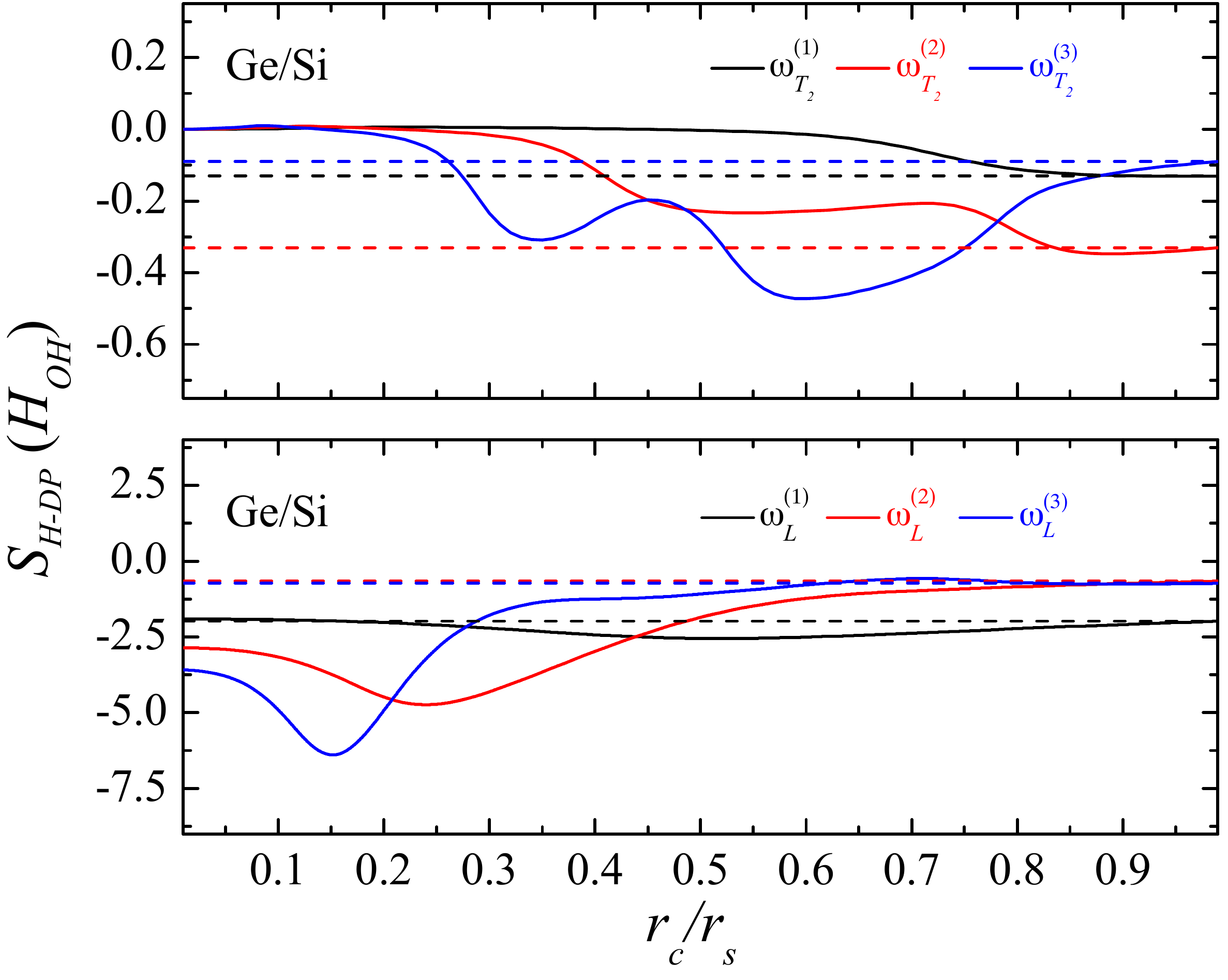}\\[0pt]
\end{center}
\caption{(Color online) Reduced valence band scattering amplitude $S_{H-BP}/H_{OH}$ for Ge/Si NWs as a function of the ratio $r_{c}/r_{s}$. For $k_{z}=0$ and $n=0$, the two sets of independent subspaces are displayed (see text): Upper panel transverse phonons with frequencies, $\omega_{T_2}^{(j)}$  ($j=1,2,3$); lower panel longitudinal modes and frequencies $\omega _{_{L}}^{(j)}$. Dashed lines: homogeneous Ge NW.}
\label{Fig7}
\end{figure}

In the size-quantum limit and not too large values of $k_{z}$, the Luttinger-Kohn (LK) Hamiltonian splits into two independent $2\times2$ matrices,
coupling ($\left\vert hh^{+}\right\rangle$, $\left\vert lh^{-}\right\rangle)$ or ($\left\vert hh^{-}\right\rangle$, $\left\vert lh^{+}\right\rangle)$
Bloch states (see Appendix~\ref{Valence-Band}). For $k_{z}=0$ and angular momentum quantum number $n=0$, the scattering amplitude~(\ref{SHHbp}) splits
into two independent terms, which correspond to the subspaces $L$ and $T_{2}$ of the hole-phonon interaction Hamiltonian.

Figure~\ref{Fig7} is devoted to the hole scattering amplitude~(\ref{SHHbp}) in units of $H_{OH}=-a(\Gamma _{15v})\sqrt{\hbar \omega _{n=0}(k_{z}=0)/4\pi
r_{c}^{2}L_{0}\rho _{c}v_{_{L_0}}^{2}}$ for the first three $T_{2}$ transverse modes (upper panel) and three $L$ longitudinal modes (lower panel) of the Ge/Si structure as a function of the ratio $r_{c}/r_{s}$. In the calculation we assumed that the lower hole state is completely confined in the core (hard wall potential approximation). As in Fig.~\ref{Fig6}, dashed lines represent the form factor for Ge NW with a radius of 5 nm. Here, the influence of the shell is solely due to Ge/Si phonon spectrum. From the figure we observe that $S_{H-BP}$ for the longitudinal modes are one order of magnitude larger than the transverse ones, reflecting the strength of coupling
between hole states. In the case of $T_{2}$ we have a coupling between $\left\vert hh\right\rangle $ and $\left\vert lh\right\rangle$, while for
the $L$ we are in the presence of the diagonal transitions $\left\vert hh\right\rangle \rightarrow \left\vert hh\right\rangle $ and $\left\vert
lh\right\rangle \rightarrow \left\vert lh\right\rangle$. Another feature is the strong oscillation of the $S_{H-BP}$ for transverse modes with
respect to the $L$ phonons. The $T_{2}$ vibrations couple the cylindrical function of second order, while for the $L$ modes, $S_{H-BP}$ is
proportional to the Bessel function $J_{0}$. In addition, a useful result can be extracted from Fig.~\ref{Fig7}, that is, we can obtain the minimum value
of $r_{c}/r_{s}$ where the hole-phonon Hamiltonian for core/shell NWs can be considered as a pure Ge wire. Notice that this result depends on the type of
interaction; for $L$ modes: $r_{c}/r_{s}\geq 0.6$ and for $T_{2}$ modes: $r_{c}/r_{s}\geq 0.8$.

\section{\label{Conclusion}Conclusion}

In summary, we have studied the acoustical phonon dispersions, the phonon displacement vectors and the electron- and hole-acoustical phonon Hamiltonians in
core/shell Ge/Si and Si/Ge NWs. Our results show the influence of the core radius and shell thickness of Si-Ge based nanowires on the phonon frequencies and electron-phonon and hole-phonon interaction Hamiltonians. Due to the presence of the shell, the phonon frequencies exhibit oscillations as function of the ratio $r_{c}/r_{s}$ leading to a strong influence on the interaction Hamiltonians and scattering amplitudes.  The gapless phonons have a tuned renormalized group sound velocities in terms of
the geometrical factor $r_{c}/r_{s}$.  Also, it is shown that scattering amplitudes for the conduction and valence bands can be handled by the shell thickness. The obtained results can be viewed as a basic tool for exploration of electron and hole transport phenomena and Brillouin light scattering, as well as for device applications of these one-dimensional Ge/Si and Si/Ge core/shell nanostructures. The systematic derivation and explicit, relatively simple, solutions of the electron and Bir-Pikus hole deformation potential Hamiltonians, incorporating the characteristics of the phonon modes for the wavenumber $k_{z}=0$, present straight applications to the resonant Raman scattering processes in core/shell NWs. Thus, searching at different light scattering configurations of the Brillouin and Raman processes, it possible to study the dependence of the $L$ and $T_{2}$ phonon modes on the spatial confinement and the intrinsic stress at the interface.

\appendix

\section{\label{Phonon} Stress tensor}

For Ge and Si semiconductors with diamond structure, the relation between stress and strain, $\mathbf{\sigma =C\cdot \varepsilon}$, in cylindrical
coordinates, $\mathbf{r}=(r,\theta ,z)$ can be written as

\begin{equation}
\left(
\begin{array}{c}
\sigma _{rr} \\
\sigma _{\theta \theta } \\
\sigma _{zz} \\
\sigma _{r\theta } \\
\sigma _{rz} \\
\sigma _{\theta z} \\
\end{array}
\right) =\left(
\begin{array}{cccccc}
C_{11} & C_{12} & C_{12} & 0 & 0 & 0 \\
C_{12} & C_{11} & C_{12} & 0 & 0 & 0 \\
C_{12} & C_{12} & C_{11} & 0 & 0 & 0 \\
0 & 0 & 0 & C_{44} & 0 & 0 \\
0 & 0 & 0 & 0 & C_{44} & 0 \\
0 & 0 & 0 & 0 & 0 & C_{44} \\
\end{array}
\right) \left(
\begin{array}{c}
\varepsilon _{rr} \\
\varepsilon _{\theta \theta } \\
\varepsilon _{zz} \\
2\varepsilon _{r\theta } \\
2\varepsilon _{rz} \\
2\varepsilon _{\theta z} \\
\end{array}
\right) ,  \label{hook_law}
\end{equation}
where the $C_{ij}$ are the elastic stiffness coefficients and the components of the strain tensor in terms of the phonon displacement vector $\mathbf{u}
=(u_{r},u_{\theta },u_{z})$ are given by

\begin{multline}
\varepsilon _{rr}=\frac{\partial u_{r}}{\partial r}\;;\text{ \ \ \ \ }
\varepsilon _{\theta \theta }=\frac{1}{r}\left( \frac{\partial u_{\theta }}{
\partial \theta }+u_{r}\right) \;; \\
\varepsilon _{zz}=\frac{\partial u_{z}}{\partial z}\;,\text{ \ }\varepsilon
_{r\theta }=\frac{1}{2}\left( \frac{1}{r}\frac{\partial u_{r}}{\partial
\theta }+\frac{\partial u_{\theta }}{\partial r}-\frac{u_{\theta }}{r}
\right) \;; \\
\varepsilon _{\theta z}=\frac{1}{2}\left( \frac{\partial u_{\theta }}{
\partial z}+\frac{1}{r}\frac{\partial u_{z}}{\partial \theta }\right) \;;
\text{ \ }\varepsilon _{rz}=\frac{1}{2}\left( \frac{\partial u_{r}}{\partial
z}+\frac{\partial u_{z}}{\partial r}\right) \;.  \label{stresscomp}
\end{multline}
Considering isotropic bulk materials the acoustic phonon branches at $\Gamma$-point are degenerate and we have that $C_{11}=\rho v_{_L}^{2}$, $C_{44}=\rho
v_{_T}^{2}$ and $C_{12}=\rho v_{_L}^{2}-2\rho v_{_T}^{2}$ with $v_{_T}(v_{_L})$ the transverse (longitudinal) sound velocity. Accordingly the stress tensor is reduced to
\begin{figure}[tbh]
\begin{center}
\includegraphics[width=\columnwidth]{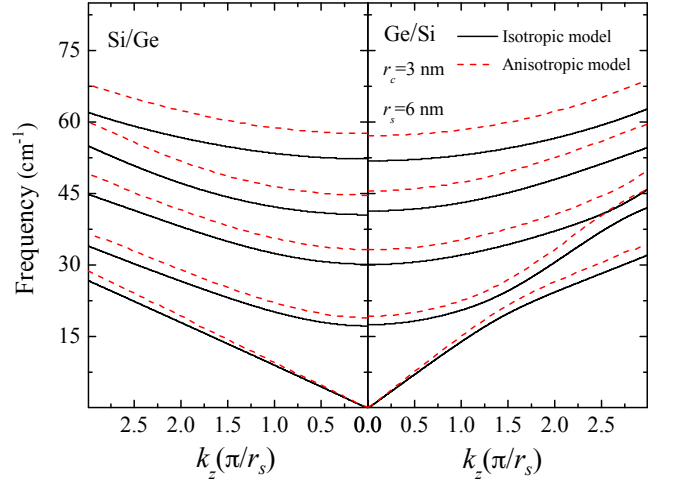}\\[0pt]
\end{center}
\caption{(Color online) $T_{2}$ phonon frequencies as a function of the phonon wavevector $k_z$ for $n=0$. Solid lines correspond to the isotropic model. Red dashed lines represent the calculations as given by Eq.~(\ref{Delta}) along [110] direction including the anisotropy effect.}
\label{Fig8}
\end{figure}
\begin{equation}
\mathbf{\sigma }=\rho (v_{_L}^{2}-2v_{_T}^{2})(\nabla \cdot \mathbf{u})
\mathbf{I}+2\rho v_{_T}^{2}(\nabla \mathbf{u})\;, \label{Isotropico}
\end{equation}
where $\mathbf{I}$ is the identity matrix. In general the vibrational modes are anisotropic along different crystallographic directions.
To carried out the effect of anisotropy we must modify the isotropic continuum model~(\ref{Isotropico}) including the complete form of the stress tensor. We can write $\mathbf{\sigma }=\mathbf{\sigma}_S+\mathbf{\sigma}^{(1)}$, where $\mathbf{\sigma}_S$ is given by Eq.~(\ref{Isotropico}) with $v^{2}_{_T}\rightarrow\overline{v^{2}_{_T}}=(v^{2}_{_{T_1}}+v^{2}_{_{T_2}})/2$ and for $\mathbf{\sigma}^{(1)}$ we have
\begin{eqnarray}
\label{stress tensor 1}
\mathbf{\sigma}^{(1)}=\rho\Delta v^{2}_{_T}\left(
\begin{array}{ccc}
           \varepsilon_{\theta\theta} & -\varepsilon_{r\theta} & -\varepsilon_{rz}  \notag \\
           -\varepsilon_{r\theta}     & \varepsilon_{rr}       & \varepsilon_{\theta z} \notag \\
           -\varepsilon_{rz}          & \varepsilon_{\theta z} & \varepsilon_{zz} \\
\end{array}
\right)
\;,
\end{eqnarray}
and $\Delta v^{2}_{_T}=v^{2}_{_{T_1}}-v^{2}_{_{T_2}}$. Taking advantage of the fact that the relation $\Delta v^{2}_{_T}/\overline{v^{2}_{_T}}=0.23$ and 0.24 for Si and Ge, respectively, we can consider the tensor $\mathbf{\sigma}^{(1)}$ as small perturbation relative to $\mathbf{\sigma}_S$. Following the procedure developed in Ref.~\onlinecite{CubaLibro} and \onlinecite{PhysRevB.50.1611} we obtain  the frequencies of the modes including the anisotropy  as $\omega^{2}=\omega^{2}_{S}+\Delta\omega^{2}$, where
\begin{eqnarray}
\label{Anisotropy_frequency}
\Delta\omega^{2}=\frac{\langle \mathbf{u}_{S}|\hat{P}|\mathbf{u}_{S}\rangle}{\langle|\mathbf{u}_{S}|^2\rangle}\;.
\label{Delta}
\end{eqnarray}
Here, $\omega_{S}$ and $\mathbf{u}_{S}$ represent the eigenfrequency and eigenvector solution of Eq.~(\ref{Emotion}) and the operator $\hat{P}$ being defined as
\begin{eqnarray}
\hat{P}&=&\rho\Delta v^{2}_{_T}\notag \\
&\times&\left(
  \begin{array}{ccc}
   k^{2}_{z}+\frac{n^{2}}{r^{2}}  & \frac{in}{r}\hat{p}^{+}_{r} & -ik_{z}\partial_{r}  \notag \\
    \frac{in}{r}\hat{p}^{-}_{r}  & -\hat{p}^{+}_{r}\partial_{r}+\frac{1}{r^{2}}-k^{2}_{z} & -\frac{nk_{z}}{r} \notag \\
    ik_{z}\hat{p}^{+}_{r} & -\frac{nk_{z}}{r} & -\hat{p}^{+}_{r}\partial_{r}-\frac{n^{2}}{r^{2}}-2k^{2}_{z} \\
  \end{array}
\right),\notag \\
\label{P Operator}
\end{eqnarray}
with $\hat{p}^{\pm}_{r}=\partial_{r}\pm\frac{1}{r}$.  Using Eq.~(\ref{Delta}) we obtain for the modes along [110] direction with $n=0$ and $k_{z}=0$ that the anisotropy does not affect the longitudinal modes $L$ and $\Delta\omega_{_L}=0$, while for the transversal $T_1$ and $T_2$ modes,
$\omega_{_{T_1,T_2}}=1.1\omega_{S}(\small{T_1,T_2})$ are shifted $10\%$ in comparison to the isotropic case for both Ge/Si and Si/Ge NWs.
Figure~\ref{Fig8} displays the effect of the anisotropy on the decoupled $T_2$ ($n = 0$) phonon spectrum for Si/Ge and Ge/Si NWs. The solid lines represent the calculation assuming the isotropic model while the anisotropic case is shown by red dashed lines. In both NWs we observe a upshift, less than $10\%$, from the $T_2$ frequency modes with the correction~(\ref{Delta}). Similar calculations can be performed for any crystallographic direction and phonon modes.

\section{\label{Conduction-Band} Electron wave function}

In the framework of the Envelope Function Approximation, the electron wave function $|\Psi _{\alpha _{e}}\rangle $ in cylindrical symmetry can be
written as
\begin{equation}
<\mathbf{r}|\Psi _{\alpha _{e}}\rangle =\frac{1}{\sqrt{2V_{c}}}
F_{m}(r)e^{i(m\theta + k_{e}z)}\;,
\label{electrwave}
\end{equation}
where $V_{c}=\pi r_{c}^{2}L_0$ is the core volume, $L_0$ the normalization length,  $m\hbar $ $(m=0,1,2...)$ and $k_{e}$ are the $z$-component of the angular momentum and electron wavenumber, respectively, and $F_{m}(r)$ the radial wave function. Considering bound states, we are in presence of two cases:~\cite{PhysRevB.77.195325,Peng-Logan-APL-2010}

a) Si/Ge NW, where the states are confined in the core. Hence, it is possible to show that
\begin{equation}
F_{m}(r)=\left\{
\begin{array}{cc}
A_{m}^{(1)}J_{m}(p_{c}r)\;; & 0\leq r\leq r_{c} \\
A_{m}^{(2)}\mathcal{Q}_{m,m}^{-}(\left\vert p_{s}\right\vert r)\;; &
r_{c}\leq r\leq r_{s}
\end{array}
\right. \;,
\label{radialE}
\end{equation}
with
\begin{equation*}
\mathcal{Q}_{m,n}^{\pm} (x)=I_{m}(x)K_{n}(\gamma x) \pm I_{n}(\gamma x)K_{m}(x)\;,
\end{equation*}
\begin{equation}
A_{m}^{(1)}=\frac{1}{2}\frac{\sqrt{\mathcal{Q}_{m+1,m}^{+}(\left\vert \tilde{p}
_{s}\right\vert )\mathcal{Q}_{m-1,m}^{+}(\left\vert \tilde{p}_{s}\right\vert )}}{
\sqrt{J_{m+1}(\tilde{p}_{c})J_{m-1}(\tilde{p}_{c})}}\frac{\tilde{p}_{s}}{
\tilde{p}_{c}}\mathcal{W}_{m}(\left\vert \tilde{p}_{s}\right\vert )\;,
\label{A1}
\end{equation}
\begin{equation}
A_{m}^{(2)}=\frac{1}{2}\mathcal{W}_{m}(\left\vert \tilde{p}_{s}\right\vert
)\;,  \label{A2}
\end{equation}
and
\begin{multline}
\mathcal{W}_{m}(\tilde{p}_{s})=\left[ \mathcal{Q}_{m+1,m}^{-}(\left\vert \tilde{p
}_{s}\right\vert )\mathcal{Q}_{m-1,m}^{-}(\left\vert \tilde{p}_{s}\right\vert
)-\right. \\
\left. \gamma ^{2}\mathcal{Q}_{m+1,m}^{-}(\gamma \left\vert \tilde{p}
_{s}\right\vert )\mathcal{Q}_{m-1,m}^{-}(\gamma \left\vert \tilde{p}
_{s}\right\vert )+\right. \\
\left. \mathcal{Q}_{m+1,m}^{+}(\left\vert \tilde{p}_{s}\right\vert )\mathcal{Q}
_{m-1,m}^{+}(\left\vert \tilde{p}_{s}\right\vert )\frac{\left\vert \tilde{p}_{s}
\right\vert ^{2}}{\tilde{p}_{c}^{2}}\right] ^{-\frac{1}{2}}\;.
\end{multline}

b) In the case of For Ge/Si core/shell, the electronic states are localized in the shell and the above equations are reduced to

\begin{equation}
F_{m}(r)=\left\{
\begin{array}{cc}
A_{m}^{(1)}I_{m}(\left\vert p_{c}\right\vert r)\;; & 0\leq r\leq
r_{c} \\
A_{m}^{(2)}\mathcal{P}_{m,m}(p_{s}r)\;; & r_{c}\leq r\leq r_{s}
\end{array}
\right. \;,
\end{equation}
with the coefficients $A_{m}^{(i)}$ $(i=1,2)$ equal to

\begin{equation}
A_{m}^{(1)}=\frac{1}{2}\frac{1}{\sqrt{I_{m+1}(\left\vert \tilde{p}
_{c}\right\vert )I_{m-1}(\left\vert \tilde{p}_{c}\right\vert )}}\mathcal{R}
_{m}(\tilde{p}_{s})\;,
\end{equation}
\begin{equation}
A_{m}^{(2)}=\frac{1}{2}\frac{1}{\sqrt{\mathcal{P}_{m+1,m}(\tilde{p}_{s})
\mathcal{P}_{m-1,m}(\tilde{p}_{s})}}\frac{\left\vert \tilde{p}
_{c}\right\vert}{\tilde{p}_{s}}\mathcal{R}_{m}(\tilde{p}_{s})\;,
\end{equation}
and

\begin{multline}
\mathcal{R}_{m}(\tilde{p}_{s})=\left[ 1-\frac{\left\vert \tilde{p}
_{c}\right\vert ^{2}}{\tilde{p}_{s}^{2}}\times \right.  \notag \\
\left. \left( 1-\frac{4}{\pi ^{2}\tilde{p}_{s}^{2}}\frac{1}{\mathcal{P}
_{m+1,m}(\tilde{p}_{s})\mathcal{P}_{m-1,m}(\tilde{p}_{s})}\right) \right] ^{-
\frac{1}{2}}\;.
\end{multline}

As stated above, $c(s)$ labels the core (shell) semiconductor and $p_{c}(p_{s})$ is related to the electron energy by the equation
\begin{equation}
\overline{E}_{e}=\Delta E_{g}^{(c,s)}+\frac{\hbar ^{2}p_{c,s}^{2}}{
2m_{t}^{(c,s)}}+\frac{\hbar ^{2}k_{e}^{2}}{2m_{l}^{(c,s)}}\;,
\label{Ee}
\end{equation}
with $\tilde{p}_{c}(\tilde{p}_{s})=p_{c}r_{c}(p_{s}r_{c})$ and $m_{l}$ ($
m_{t})$ the longitudinal (transverse) conduction electron mass at $
\Gamma$-point of the Brillouin zone.~\cite{Note1}
In Eq.~(\ref{Ee}) $\overline{E}_{e}=E_{g}^{(c,s)}-E_{strained}$
takes into account the gap energy correction due to the intrinsic strain at the interface~\cite{PhysRev.101.944,JAP_101_104503_2007} and $\Delta E_{g}^{(c,s)}$ is the band offset between the core and shell measured from the bottom of the band. For NWs along [110] growth direction, the band gap $\Delta E_{g}^{(c,s)}\simeq 300$ meV.~\cite{PhysRevB.77.195325} In our calculations we have assumed $\Delta E_{g}^{(c,s)}$ independent of $\gamma$.

There is a third option, not considered here, where both, $p_{c}$ and $p_{s}$ are real, and the radial wave function $F_{m}(r)$ presents an oscillatory behavior in both the core and shell parts, which correspond to higher excited states of the core/shell NWs.

\section{\label{Valence-Band} Hole wave function}

For a description of the hole states in the valence band we consider the LK Hamiltonian model neglecting the coupling from the split-off band. This Hamiltonian provides a good description for heavy-hole and light-hole states and the coupling between them due to $\Gamma_{15v}$ degeneracy of valence bands at $\Gamma$-point. Along the [110] direction and assuming the axial approximation, $\gamma_{2}$ $\simeq $ $\gamma _{3}$, the $4 \times 4$ Hamiltonian can be written as~\cite{PhysRev.97.869,PhysRev.102.1030,Rossler1984}

\begin{equation}
H_{LK}=\frac{\hbar ^{2}}{m_{0}}\left(
\begin{array}{cccc}
D_{hh} & A_{-} & B_{-} & 0 \\
A_{-}^{\ast } & D_{lh} & 0 & B_{-} \\
B_{-}^{\ast } & 0 & D_{lh} & A_{+}^{\ast } \\
0 & B_{-}^{\ast } & A_{+} & D_{hh} \\
\end{array}
\right) \;,  \label{Hamiltoniano-LK}
\end{equation}
where

\begin{eqnarray}
D_{hh} &=&-\frac{(\gamma _{1}+\gamma _{s})}{2}\{\hat{k}_{+},\hat{k}_{-}\}-
\frac{(\gamma _{1}-2\gamma _{s})}{2}\hat{k}_{h}^{2}\;,  \label{ME} \\
D_{lh} &=&-\frac{(\gamma _{1}-\gamma _{s})}{2}\{\hat{k}_{+},\hat{k}_{-}\}-
\frac{(\gamma _{1}+2\gamma _{s})}{2}\hat{k}_{h}^{2}\;,  \notag \\
A_{\pm } &=&\mp \sqrt{3}\hat{\gamma} \hat{k}_{\pm }\hat{k}_{h}\;\text{; }B_{\pm }=-
\frac{\sqrt{3}}{2}\gamma _{t}\hat{k}_{\pm }^{2}\;,  \notag
\end{eqnarray}
$\hat{\gamma} =(\gamma _{2}+\gamma _{3})/2$, $\gamma _{s}=(\gamma _{2}+3\gamma_{3})/4$, $\gamma _{t}=(3\gamma _{2}+5\gamma _{3})/8$, $\gamma _{1}$, $\gamma _{2}$ and $\gamma _{3}$ are the Luttinger parameters. The total Hamiltonian for the valence band can be cast as $H=H_{LK}+V(r)$ with $V(r)$ the NWs confinement potential. The wave function $<\mathbf{r}|\Psi _{\alpha_{e}}\rangle$, as given by Eq.~(\ref{electrwave}), represents a basis for the effective $4\times 4$ LK-Hamiltonian. Since the Bloch states, $\left\vert hh^{+}\right\rangle$, $\left\vert lh^{+}\right\rangle$, $\left\vert lh^{-}\right\rangle$ and $\left\vert hh^{-}\right\rangle $ are mixed by the effects of the cylindrical symmetry and the non-zero matrix elements $A_{\pm}$ and $B_{\pm }$ in Eq.~(\ref{ME}), we can write the general solution of the wave function $<\mathbf{r}|\Psi _{\alpha _{h}}\rangle $ with a special sequence of the angular quantum number $m$ for each hole state. Thus, by exploring the symmetry of the Hamiltonian~(\ref{Hamiltoniano-LK}), the exact
wave function for the hole state $<\mathbf{r}|\Psi _{\alpha _{h}}\rangle $ can be written as

\begin{multline}
<\mathbf{r}|\Psi _{\alpha _{h}}^{(i)}\rangle =\widehat{F}_{m}^{(i)}(r)e^{i(m
\theta +k_{h}z)}=  \label{basic-hole} \\
\left(
\begin{array}{c}
a_{1i}F_{m}(p_{hh}r)|hh^{+}\rangle \\
a_{2i}F_{m+1}(p_{hl}r)e^{i\theta }|lh^{+}\rangle \\
a_{3i}F_{m+2}(p_{hl}r)e^{2i\theta }|lh^{-}\rangle \\
a_{4i}F_{m+3}(p_{hh}r)e^{3i\theta }|hh^{-}\rangle
\end{array}
\right) e^{i(m\theta +k_{h}z)}\;,
\end{multline}
where $p_{_{hh(lh)}}$ is related to the heavy (light) hole energy by the expression

\begin{equation}
\overline{E}_{hh(lh)}=-\Delta E_{g}^{(c,s)}-\frac{\hbar ^{2}}{2m_{hh(lh)}}(p_{hh(lh)}^{2}+k_{h}^{2})\;,
\label{EHH}
\end{equation}
and $m_{hh(lh)}=1/(\gamma _{1}-(+)2\gamma _{s})$. As in the case of the conduction band\ in $E_{hh(lh)}$ we consider the band gap correction. The vector coefficients $\mathbf{a}_{i}$ $\left\vert i\right\rangle $ ($
i=hh^{+},lh^{+},lh^{-},hh^{-})$ in~(\ref{basic-hole}) are~\cite{PhysRevB.42.3690}

\begin{eqnarray*}
\mathbf{a}_{hh^{+}}^{\mathbf{\dagger }} &=&a_{hh^{+}}\left( -\frac{1}{\sqrt{3
}}\left( 1+\frac{4k_{h}^{2}}{p_{hh}^{2}}\right) ,-\frac{2k_{h}}{p_{hh}}
,1,0\right) \;, \\
\mathbf{a}_{lh^{+}}^{\mathbf{\dagger }} &=&a_{lh^{+}}\left( -\sqrt{3},-\frac{
2k_{h}}{p_{lh}},1,0\right) \;, \\
\mathbf{a}_{lh^{-}}^{\mathbf{\dagger }} &=&a_{lh^{-}}\left( -\frac{2k_{h}}{
p_{lh}},\frac{1}{\sqrt{3}}\left( 1+\frac{4k_{h}^{2}}{p_{lh}^{2}}\right)
,0,1\right) \;, \\
\mathbf{a}_{hh^{-}}^{\mathbf{\dagger }} &=&a_{hh^{-}}\left( -\frac{2k_{h}}{
p_{hh}},-\sqrt{3},0,1\right) \;,
\end{eqnarray*}
where the wight coefficients $a_{hh^{+}}$, $a_{lh^{+}}$, $a_{lh^{-}}$, $
a_{hh^{-}}$ give a measure of the mixtures of Bloch states $\left\vert
i\right\rangle =\left\vert hh^{+}\right\rangle$, $\left\vert
lh^{+}\right\rangle$, $\left\vert lh^{-}\right\rangle $ and $\left\vert
hh^{-}\right\rangle$. Imposing continuity of the wave function $<\mathbf{r}
|\Psi _{\alpha _{h}}^{(i)}\rangle $ and its derivative at the core/shell interface $r=r_{c}$ and choosing the boundary condition $<\mathbf{r}|\Psi_{\alpha _{h}}^{(i)}\rangle \mid _{r=r_{s}}=0$, we find the normalized eigensolutions and eigenenergies for the hole states.

In the case of Ge/Si core/shell NWs, the hole are mostly confined in the core and the valence band offset is of the order 0.5 eV.~\cite{PhysRevB.77.195325,Peng-Logan-APL-2010,doi:10.1021/nl9029972} Thus, in the limit of strong spatial confinement we can assume a hard wall potential and the holes are completely confined in the core.

In the evaluation of the hole energy and wave function we employed for Si[Ge] the values $\gamma_{1}=4.22[13.4]$, $\gamma _{2}=0.39[4.24]$, $\gamma
_{3}=1.44[5.69]$, $a(\Gamma _{15v})=-5.0[-5.2]$ eV and $b(\Gamma_{8v})=-2.3[-2.4]$ eV.~\cite{Adachi}

\acknowledgments

C. T-G and G. E. M acknowledge support from the Brazilian Agencies FAPESP (proceses: 2015/23619-1, 2014/19142-2) and CNPq. V. Romero is acknowledged for critical reading of the manuscript. D.S-P wishes to thank the CNPq-CLAF program for financial support.

\bibliographystyle{apsrev}
\bibliography{ReferenceMarzo}

\end{document}